\documentclass[11pt, executivepaper]{article}
\usepackage[utf8]{inputenc}
\usepackage[T1]{fontenc}
\usepackage{natbib}
\usepackage{amsmath}
\usepackage{amssymb}
\usepackage{xcolor}
\usepackage{amsfonts}
\usepackage{graphicx}
\usepackage{enumitem}
\usepackage{geometry}
\setcitestyle{aysep={}}
\begin{document}

\title{\textbf{A No-Go Theorem for $\psi$-ontic Models? Yes! Response to Criticisms}}

\author{Gabriele Carcassi\thanks{Physics Department, University of Michigan, Ann Arbor, MI 48109. E-mail: carcassi@umich.edu} \and Andrea Oldofredi\thanks{Centre of Philosophy (Lancog), University of Lisbon, Portugal. E-mail: aoldofredi@letras.ulisboa.pt} \and Christine A. Aidala\thanks{Physics Department, University of Michigan, Ann Arbor, MI 48109. E-mail: caidala@umich.edu}}

\maketitle

\begin{abstract}
This short note addresses the criticisms recently proposed by Shan Gao against our article ``{\em On the Reality of the Quantum State Once Again: A No-Go Theorem for $\psi$-Ontic Models}'' (Found.\ Phys.\ 54:14). The essay aims to respond to such objections and to show once again that the theorem proved in our paper is correct, and therefore true---contrary to Gao's claims. Philosophical consequences of this fact are briefly discussed.
\vspace{5mm}

\end{abstract}
\vspace{5mm}

\begin{center}
Accepted for publication in {\em Foundations of Physics}
\end{center}

\vspace{5mm}

\tableofcontents

\newpage

\section{Gao's Criticisms Against Our No-Go Theorem for $\psi$-Ontic Models}
\label{Intro}

Earlier this year, we published the paper titled ``{\em On the Reality of the Quantum State Once Again: A No-Go Theorem for $\psi$-Ontic Models}'' (\cite{Carcassi:2024}) in Foundations of Physics, demonstrating that $\psi$-ontic models as defined by \cite{Harrigan:2010} (HS) cannot reproduce all the results and consequences of quantum mechanics (QM). The no-go theorem contained in our essay gathered the attention of some experts in quantum foundations and one of them, Shan Gao, provided a criticism against our result in an article that recently appeared in the same journal (\cite{Gao:2024}). 

Since Gao himself notes that if our argument is correct, ``it will be a very important new result'', in the present work we respond to his objection showing that our theorem is indeed true, contrary to the claims contained in his paper. Referring to this, we will include a final discussion highlighting the important philosophical consequences of our theorem.


Before delving into our response, let us briefly summarize the argument included in \cite{Gao:2024}. This paper firstly reviews the HS ontological model framework, and then the technical result of the entropy calculation from our article. Gao proceeds to analyze the difference between von Neumann and Shannon entropy, noting that the von Neumann entropy is the Shannon entropy calculated on orthogonal states. Therefore, he says, using the Shannon entropy on non-orthogonal states yields wrong results---i.e.\ results contradicting quantum theory---because it assumes that such states are as distinguishable as orthogonal ones. In the words of the author, the error we made in the derivation of our theorem can be stated as follows:

\begin{quote}
The issue lies in that COA [Carcassi, Oldofredi, Aidala] implicitly assumed that
in $\psi$-ontic models, the two ontic states $\lambda_\psi$ and $\lambda_\phi$, which are represented by two non-orthogonal states $|\psi\rangle$ and $|\phi\rangle$, are classical states that can be distinguished by experiments with certainty (\cite{Gao:2024}, p.\ 4).
\end{quote}

\noindent Thus, according to Gao our result crucially relies on an implicit assumption that is {\em not} contained in the original Harrigan and Spekkens' model, i.e.\ that any pair of states can be distinguished by experiments. In Gao's opinion, then, what we proved

\begin{quote}
is only that these models with an additional wrong assumption of the distinguishability of non-orthogonal states are inconsistent with QM ({\em ibid.}, p.\ 5).
\end{quote}

In order to respond to this objection, let us ask the question whether we did or did not employ such an additional assumption in our paper. Moreover, in our article we laid out a series of issues one would have to solve to be able to fix the problems we identified for the HS framework. How does Gao's paper address those issues? 

\section{Did we use an additional assumption?}
\label{2}

The fact that the Shannon/Gibbs entropy, unlike the von Neumann entropy, implicitly assumes that all states are equally distinguishable is exactly the feature we use to reach our result. There is no doubt about that. Gao's paper, however, inappropriately left out the motivation as to why we used the Shannon/Gibbs entropy in the first place. Without stating such a motivation, the understanding of our theorem---as well as its derivation---is simply compromised. In order to clarify this crucial point to the reader, let us go through our reasoning once again.

As Gao writes when reviewing the ontological model framework of Harrigan and Spekkens,
\begin{quote}
a wave function or a pure state corresponds to a probability distribution $p(\lambda| P)$ over all possible ontic states when the preparation is known to be $P$ ({\em ibid.}, p.\ 2).

\end{quote}

\noindent In addition, 

\begin{quote}
a mixture of pure states $|\psi_i\rangle$ with probabilities $p_i$ is represented by $\sum_i	p_i p(\lambda| P_{\psi_i})$ ({\em ibid.}).

\end{quote}
That is, from these quotes it is clear that \emph{the model itself assumes that a quantum statistical mixture is a simply a classical probability distribution over pure states, which are themselves classical probability distributions over the ontic states}.
This is not an implicit assumption that we made; rather, it is an assumption made by Harrigan and Spekkens. This is the problematic aspect of their model which renders it too classical, as we stated multiple times in the paper, including in our conclusion:
\begin{quote}
The key problem of the HS categorization is that it is “not quantum enough”. [...] HS definitions entail that statistical mixtures are simply classical
mixtures in the sense that they are captured by a standard probability measure (\cite{Carcassi:2024}, p.\ 11).
\end{quote}

As it is well-known, classical probability already assumes that all elements are equally distinguishable. Suppose that $(\Lambda, \Sigma_{\Lambda}, p_P)$ is a probability space consisting of the space of ontic states $\Lambda$, its $\sigma$-algebra $\Sigma_{\Lambda}$, and the probability measure $p_P$ corresponding to some generic preparation $P$. Let's also pick two ontic states $\lambda_1, \lambda_2 \in \Lambda$ and define $f(\lambda)=\textbf{1}_{\lambda_1} - \textbf{1}_{\lambda_2}$, where $\textbf{1}$ is the indicator function. This is a random variable that takes the value $1$ over $\lambda_1$, $-1$ over $\lambda_2$ and 0 otherwise. That is, it is a random variable that fully distinguishes between the two states. 
In turn, this is exactly why a Kolmogorov probability space cannot replicate the results of quantum mechanics: it can fully distinguish between everything. If one wants to be able to assign a probability (or a probability density) to each ontic state, the $\sigma$-algebra has to be fine enough to include all ontic states. This is the mathematical problem at stake with the HS framework. And it is not a new problem: the fact that quantum mechanics is not a Kolmogorovian probability theory has been known since the very inception of the theory.

Now, what does the ontological model framework do that is slightly new? We say that in our article multiple times, including the conclusion:
\begin{quote}
[t]he framework in fact tries to hide the non-classicality at the level of each ontic state which, by construction, cannot be decomposed further using a standard probability measure ({\em ibid.}). 
\end{quote}

\noindent Harrigan and Spekkens do this by having all measurements be different events (i.e.\ things one can condition on, elements of the $\sigma$-algebra) instead of random variables. Effectively, the sample space of the probability space is not $\Lambda$, as in our example immediately above, but $\Lambda \times \mathcal{M}$, where $\mathcal{M}$ is the space of all possible measurements. This allows one to set the correct probability of quantum mechanics for pure states, but it leaves the door open to problems for mixed states.

Having made it clear that no implicit assumption of classicality has been made in our paper, let us see how our argument proceeded. Since we reached the conclusion that, in the framework of the ontological model, statistical mixtures are modeled as classical probability distributions (i.e.~probability measures), we should expect to use the tools of classical statistical mechanics. Why? Well, simply because we only have two successful theories of statistical mechanics: the classical one that works on Kolmogorovian probability measures, and the quantum one that works on density operators. In our article, we mention multiple times that the assumption that epistemic states mix like classical probability is critical. For example:

\begin{quote}
Given that $M(\Lambda)$ is the space of probability measures over $\Lambda$, it is natural to use the Shannon/Gibbs entropy function $H_\Lambda : M(\Lambda) \to \mathbb{R}$, as this is what is typically done in statistical mechanics ({\em ibid.}, p.\ 5).
\end{quote}

\noindent That is, we use the Shannon/Gibbs entropy because this is the entropy connecting to thermodynamics when working on classical distributions. 

In short, our full argument can be schematized as follows:
\begin{enumerate}
	\item The ontological model assumes that quantum statistical mixtures are (classical) probability distributions (i.e.~probability measures) over the space of ontic states $\Lambda$.
	\item To fully replicate quantum mechanics, an ontological model must replicate {\em all} its results, including those from quantum statistical mechanics, quantum thermodynamics and quantum information theory.
	\item Therefore it must, at least, compute entropies correctly.
	\item Given that mixtures in the HS model framework are modeled as classical probability distributions, we should be entitled to use the only tools we have that work on those (i.e.~classical statistical mechanics).
	\item The Shannon/Gibbs entropy of classical statistical mechanics cannot replicate the von Neumann entropy.
	\item $\therefore$ Premises 1.--5. entail that the HS framework has a problem. 
\end{enumerate}
The issue with Gao's paper is that it summarizes our argument starting at point 5, without mentioning all previous essential elements of the argument. In particular, the author charges our derivation with an implicit assumption of classicality  without appreciating the fact that such an assumption in embedded in the original HS model, as made explicit in premise 1.
Without context, we fully agree that point 5 by itself is not enough to proceed to the conclusion stated in 6. But there is significant additional context provided in our paper, which Gao's reply does not mention. 

In conclusion, let us repeat again that we did not employ any implicit premise. Contrary to Gao's belief, such an assumption is already embedded into the HS model when it requires that epistemic states combine classically, as we stated multiple times in our article.

\section{{Did Gao address the outstanding issues?}}
\label{3}
Setting aside the specific points above, there remain other issues with the ontological model framework that Gao's paper does not address or acknowledge. As we say in the conclusion of our original paper, an ontological model must be 

\begin{quote}
able to reproduce all results from quantum statistical	mechanics (e.g.\ derive the Boltzmann distribution from entropy maximization) or quantum information theory
(e.g.\ derive the Holevo bound). Since these are derived from the more general framework of quantum mechanics, one cannot cherry pick which results a model has
to reproduce to be a valid model of quantum theory: all results must be correctly reproduced ({\em ibid.}, p.\ 11).
\end{quote}

\noindent Gao does not seem to disagree. We will assume that we all agree on this point, because if we do not the whole argument is already moot.

How are we supposed to proceed? How are we supposed to recover all results of, say, quantum statistical mechanics? Gao's paper is not explicit about this. The only comment is in regard to the entropy:

\begin{quote}
thus the information entropy of a mixture of these states should be given not by the Shannon entropy, but by the von Neumann entropy (when further considering the fact that two mixed states that have the same density matrix cannot be distinguished by experiments) (\cite{Gao:2024}, p.\ 5).
\end{quote}

\noindent It should be noted here, however, the von Neumann entropy cannot be calculated on an epistemic state, which is a Kolmogorov probability measure: the von Neumann entropy requires a density matrix. Is this not simply an admission that one cannot calculate it directly on the epistemic state? Suppose we gave someone only the ontological model, without knowledge of Hilbert spaces. Are we saying that the individual would not be able to calculate the entropy? If that is the case, we cannot claim that the ontological model fully reproduces quantum mechanics. In our view, saying ``Use the von Neumann entropy'', as Gao seems to suggest, is precisely saying``We cannot replicate this result in the ontological model framework,'' thus confirming our no-go result under scrutiny.

Note that, in section 4 of our original paper, we do leave the door open for a possible alternative theory:
\begin{quote}
That is, one may give a different definition of entropy to use on the ontic space (\cite{Carcassi:2024}, p.\ 7).
\end{quote}
We then list the two biggest problems one would face in attempting to construct a suitable alternative definition of entropy; there are additional technical problems we do not even mention. We are open to the possibility that there may be something that can be made to work, though we are skeptical that this can be done with the simple ``quantum state at time $t$ $\leftrightarrow$ ontic state at time $t$'' relationship assumed by the HS categorization. Gao's paper does not mention those problems, and it does not even show how one can recover the density matrix from an epistemic state so that the von Neumann entropy can be calculated.

Referring to this, it must be underlined that the entropy is just one element of quantum statistical mechanics. Consider for instance a spin-1/2 system. If it is prepared in the $z^+$ direction, the state will have a vertical rotational symmetry. Similarly, if it is prepared in the $x^+$ direction it will have a horizontal rotational symmetry. Now suppose we have an equal mixture of $z^+$ and $x^+$. This will have a rotational symmetry along the diagonal between the two. We can see it geometrically by considering the Bloch ball. The mixture will be the midpoint between the topmost point and the rightmost point. This midpoint will be on the 45-degree axis, and therefore a rotation along that axis will leave the mixture unchanged. Mathematically, we can simply find the eigenstates of the density matrix and find a unitary operator over the same eigenstates, or, equivalently, any unitary operator that commutes with the density matrix. How can this reasoning be replicated within an ontological model? The corresponding epistemic state of the mixture does not have that symmetry, so where does it come from? What are the eigenstates of a generic epistemic state? None of this is answered in Gao's paper.

The overall question is: how does one do quantum statistical mechanics, quantum thermodynamics and quantum information theory within an ontological model? Gao's paper does not address these issues at all: the term ``statistical mechanics'' does not even appear in his essay. Naturally, we can go back to the Hilbert space and calculate everything there, but that is also true for the probability of transition $p(k | \lambda,  M)$. Can we really consider a model successful if it can provide only some of the results of quantum theory (i.e.~expectations) but not the rest? Our position is no. Quantum statistical mechanics is not simply classical statistical mechanics on top of quantum pure states.

\section{Discussion and Conclusion}

While Gao's paper claims that we allegedly used an implicit premise, in this note we clarified that such an assumption is already embedded into Harrigan and Spekkens' use of classical probability by employing classical mixing of epistemic states. This fact was emphasized several times in our original article, but unfortunately it is not acknowledged in Gao's work. Therefore, his criticism to our theorem for $\psi$-ontic models is rejected.\footnote{We would like to emphasize that the HS model itself does makes use of several implicit assumptions concerning the nature of the quantum state, as underlined in \cite{Oldofredi:2020} and \cite{Oldofredi:2021}. However, Gao does not mention such implicit assumptions used by Harrigan and Spekkens in his essay.} Moreover, our article highlighted problems in recovering quantum statistical mechanics from the ontological model framework. We do not see that Gao addressed, or in fact even mentioned, those issues, even though they were raised by Carcassi in a signed response as part of the peer-review process. 

In conclusion, let us address two more points which we think have philosophical significance and that were not properly stated in \cite{Carcassi:2024}. In the first place, no-go theorems in physics are \emph{never} ruling out ``everything''. They always rule out something defined \emph{in a particular way}. For instance, the Bell-Kochen-Specker theorem does not rule out {\em all} hidden variables theories, only local ones, the Pusey-Barrett-Rudolph theorem (PBR) does not rule out {\em all} epistemic models, but only those falling into the HS framework (cf.\ \cite{Oldofredi:2020}, \cite{Oldofredi:2021} for details). Similarly, our no-go result does not rule out {\em all} ontic models, i.e.\ it does not refute all those models and theories aiming at providing an ontic---or better, realist---understanding of the quantum state, but only those ones following the operational definitions of the HS framework. We would in fact welcome an ontological model that replicates \emph{all} of QM, including quantum statistical mechanics. Referring to this, let us consider for the sake of the discussion Bohmian mechanics, a well-known realist interpretation of the quantum formalism. In our opinion it is not trivial to say that such a theory conforms to the definitions given by Harrigan and Spekkens. On the one hand, the HS framework is framed within operational QM, and its basic ingredients are (i) inherent properties of individual physical systems associated with experimental protocols, and (ii) measurements performed on such systems revealing the value of some such particular properties. It is very well-known, however, that according to Bohmian mechanics quantum observables are {\em not} genuine attributes of the corpuscles, and that quantum measurements do {\em not} reveal the values of such ``intrinsic'' properties (\cite{Durr:2013}, Chapter 3). Thus, the very assumptions made by the ontological model framework seem at odds with respect to the metaphysical tenets of the Bohmian theory. On the other hand, in $\psi$-ontic models the quantum state is supposed to be an entity representing physical reality either completely or partially (\cite{Harrigan:2010}, p.\ 129). However, in most contemporary readings of Bohmian mechanics the quantum wave function is not considered a real physical entity---i.e.\ it is not part of the three-dimensional beables defined in space-time as required by the HS approach---but it is instead defined as a nomological, abstract object that guides the dynamical behavior of the particles (\cite{Durr:2013}, \cite{Esfeld:2017}). For these motivations, we believe that one has to show how the Bohmian theory---if taken seriously---fits into the HS framework, until then we suspend the judgement as to whether our theorem applies to this interpretation, showing again that not every realist understanding of QM is ruled out by our result. 

The related second point concerns the interpretation of the PBR theorem. In \cite{Carcassi:2024} we suggested that such no-go theorem must be understood as an argument aimed at showing a problem for the HS model framework, since the latter is a crucial assumption of the former. More precisely, we claimed that the PBR theorem finds that $\psi$-epistemic models are untenable
because the HS categorization employs standard probability theory for composite systems, as we underlined in Section \ref{2} of the present text.
We furthermore suggested that the logical conjunction of PBR and our result implies the emptiness of Harrigan and Spekkens' model:
\begin{align*}
\textrm{PBR theorem} \wedge \textrm{COA theorem} \Longrightarrow \neg \textrm{HS framework},
\end{align*}

\noindent given that both $\psi$-ontic and $\psi$-epistemic models are shown to contradict the predictions and results of quantum mechanics. This is why we argued that the HS categorization should not be used in order to evaluate the nature of the quantum state. 

However, we are well aware that the general understanding of the PBR theorem is not what we suggested. Rather, experts in quantum foundations as well as philosophers of physics usually claim that the PBR argument entails an ontological reading of the quantum wave function, since it excludes $\psi$-epistemic models (cf.\ for instance \cite{BrancIard:2014}, \cite{Maudlin:2019}, \cite{Reich:2011}, \cite{Wallace:2020}). We disagree with this received view for two main reasons:

\begin{enumerate}
\item As just stated a few lines above, a no-go theorem does not exclude ``everything'': even assuming that the PBR theorem excludes $\psi$-epistemic models as defined by HS, it does not refute the entire class of epistemic interpretations of quantum theory. Notable examples of epistemic quantum theories not rejected by PBR are Entropic Dynamics,\footnote{Note that before eq.\ 27 of \cite{Caticha:2022} the author identifies the same problematic issue in the HS categorization.} Relational quantum mechanics, Quantum Bayesianism and Perspectival quantum mechanics to mention a few (cf.\ \cite{Ben-Menahem:2017} for other examples). Thus, it should be stated more correctly that the PBR theorem only applies to those particular theories satisfying the HS definition of $\psi$-epistemic model, which is just a subset of all epistemic readings of the quantum state.

\item Since in our previous article we showed that the HS model framework is fundamentally flawed in the sense that it contains a premise that is incompatible with quantum mechanics, we believe that no substantial metaphysical readings of the PBR argument should be given. Logically, if the HS model framework is shown to be flawed, then any other formal result derived from it will suffer from the very same problem affecting the original approach. Consequently, the PBR theorem should be used neither as an argument in favor, nor against particular interpretations of the quantum state. In our opinion, the PBR result does show that $\psi$-epistemic models as defined by HS contradict quantum theory, and consequently, it tells us that we should not use such ontological model framework in order to evaluate epistemic interpretations of the quantum formalism. A similar conclusion is obtained in our no-go theorem for $\psi$-ontic models.
\end{enumerate}

In summary, in this essay we have expressed the reason why we do not believe that any of the criticisms leveled at our previous work~\cite{Carcassi:2024} are valid, and clarified what we think is the correct interpretation of the PBR theorem.

\vspace{5mm}

\textbf{Acknowledgements:} AO acknowledges financial support for this research from the Fundacao para a Ciencia e a Tecnologia (FCT) (Grant no. 2023.07796.CEECIND).

This paper is part of the ongoing \textit{Assumptions of Physics} project \cite{aop-book}, which aims to identify a handful of physical principles from which the basic laws can be rigorously derived.
\clearpage

\bibliographystyle{apalike}
\bibliography{bibliography}

\end{document}